\documentclass[10pt,conference,anonymous]{IEEEtran}
\IEEEoverridecommandlockouts

\usepackage{amsmath,amsfonts, amssymb}
\usepackage{algorithm}
\usepackage{adjustbox}
\usepackage{lscape}
\usepackage{siunitx}
\usepackage[noend]{algpseudocode}
\usepackage{textcomp}
\usepackage{fancyvrb}
\usepackage{graphicx}
\usepackage{soul}
\usepackage[many]{tcolorbox}    	% for COLORED BOXES (tikz()
\usepackage{setspace}      
\usepackage{booktabs}
\usepackage[utf8]{inputenc}
\usepackage[T1]{fontenc}
\usepackage{rotating}
\usepackage{graphicx}
\usepackage{paralist}
\usepackage{tabularx}
\usepackage{multicol}
\usepackage{multirow}
\usepackage{pbox}
\usepackage{enumitem}	
\usepackage{colortbl}
\usepackage{pifont}
\usepackage{xspace}
\usepackage{url}
\usepackage{tikz}
\usepackage{tabularx}
\usepackage{fontawesome}
\usepackage{lscape}
\usepackage{listings}
\usepackage{color}
\usepackage{showexpl}
\usepackage{anyfontsize}
\usepackage{comment}
\usepackage{soul}
\usepackage{multibib}
\usepackage{multirow}
\usepackage{xcolor}
\usepackage{xspace}
\usepackage{caption}
\usepackage{footnote}
\usepackage{tcolorbox}
\usepackage{booktabs}
\usepackage{fixltx2e}
\usepackage{balance}
\usepackage[normalem]{ulem}

\usepackage{hhline}
\definecolor{main}{HTML}{5989cf}    % setting main color to be used
\definecolor{sub}{HTML}{cde4ff}     % setting sub color to be used

\tcbset{
	sharp corners,
	colback = white,
	before skip = 0.2cm,    % add extra space before the box
	after skip = 0.5cm      % add extra space after the box
}                           % setting global options for tcolorbox

\newtcolorbox{boxM}{
	fontupper = \color{white},
	rounded corners,
	arc = 6pt,
	colback = main!80, 
	colframe = main, 
	boxrule = 0pt, 
	bottomrule = 4.5pt,
	enhanced,
	fuzzy shadow = {0pt}{-3pt}{-0.5pt}{0.5pt}{black!35}
}
\newcolumntype{N}{>{\centering\arraybackslash}m{.85in}}
\def\BibTeX{{\rm B\kern-.05em{\sc i\kern-.025em b}\kern-.08em
    T\kern-.1667em\lower.7ex\hbox{E}\kern-.125emX}}

\newboolean{showcomments}

\setboolean{showcomments}{true}

\ifthenelse{\boolean{showcomments}}
{\newcommand{\nb}[2]{
		\fbox{\bfseries\sffamily\scriptsize#1}
		{\sf\small$\blacktriangleright$\textit{#2}$\blacktriangleleft$}
	}
	
}
{\newcommand{\nb}[2]{}
	
}

\newcommand{\satd}{\emph{Self-Admitted Technical Debts}\xspace}
\newcommand{\cmark}{\ding{51}}%
\newcommand{\xmark}{\ding{55}}%
\newcommand{\ie}{\emph{i.e.,}\xspace}
\newcommand{\eg}{\emph{e.g.,}\xspace}
\newcommand{\etc}{etc.\xspace}
\newcommand{\etal}{\emph{et~al.}\xspace}
\newcommand{\secref}[1]{Section~\ref{#1}\xspace}
\newcommand{\figref}[1]{Fig.~\ref{#1}\xspace}

\newcommand{\tabref}[1]{Table~\ref{#1}\xspace}

\newcommand{\java}{\emph{Java}\xspace}
\newcommand{\tool}{\emph{SATDBailiff}\xspace}

\newcommand*\circled[1]{\tikz[baseline=(char.base)]{
		\node[shape=circle,fill,inner sep=0.8pt] (char) {\textcolor{white}{#1}};}}

\definecolor{lightergray}{rgb}{0.9,0.9,0.9}
\newtcolorbox{resultbox}{colback=lightergray, arc=0.5mm, top=2mm, bottom=2mm, left=2mm, right=2mm}

\newcommand\rev[1]{\textcolor{black}{#1}}
\begin{document}

\title{Towards Automatically Addressing Self-Admitted Technical Debt: How Far Are We?}

\author{
	\IEEEauthorblockN{Antonio Mastropaolo}
	\IEEEauthorblockA{\textit{SEART @ Software Institute,}\\
		\textit{Universit\`{a} della Svizzera italiana (USI)},\\ Switzerland}
	 \and 
	 \IEEEauthorblockN{Massimiliano Di Penta}
	 \IEEEauthorblockA{\textit{Dept. of Engineering,}\\ \textit{University of Sannio,}\\ Italy}
	 \and
	 \IEEEauthorblockN{Gabriele Bavota}
\IEEEauthorblockA{\textit{SEART @ Software Institute,}\\
	\textit{Universit\`{a} della Svizzera italiana (USI)},\\ Switzerland}
	%\and	
	%\IEEEauthorblockA{\IEEEauthorrefmark{1}\textit{SEART @ Software Institute, Universit\`{a} della Svizzera italiana (USI), Switzerland}}
	%\IEEEauthorblockA{\IEEEauthorrefmark{2}\textit{University of Sannio, Italy}}

}

\maketitle

\thispagestyle{empty}

\begin{abstract}
	Upon evolving their software, organizations and individual developers have to spend a substantial effort to pay back technical debt, \ie the fact that software is released in a shape not as good as it should be, \eg in terms of functionality, reliability, or maintainability.
	This paper empirically investigates the extent to which technical debt can be automatically paid back by neural-based generative models, and in particular models exploiting different strategies for pre-training and fine-tuning. 
	We start by extracting a dateset of 5,039 Self-Admitted Technical Debt (SATD) removals from 595 open-source projects. SATD refers to technical debt instances documented (\eg via code comments) by developers. We use this dataset to experiment with seven different generative deep learning (DL) model configurations. Specifically, we compare transformers pre-trained and fine-tuned with different combinations of training objectives, including the fixing of generic code changes, SATD removals, and SATD-comment prompt tuning. Also, we investigate the applicability in this context of a recently-available Large Language Model (LLM)-based chat bot. 
	Results of our study indicate that the automated repayment of SATD is a challenging task, with the best model we experimented with able to automatically fix $\sim$2\% to 8\% of test instances, depending on the number of attempts it is allowed to make. Given the limited size of the fine-tuning dataset ($\sim$5k instances), the model's pre-training plays a fundamental role in boosting performance. Also, the ability to remove SATD steadily drops if the comment documenting the SATD is not provided as input to the model. Finally, we found general-purpose LLMs to not be a competitive approach for addressing SATD. 
\end{abstract}


\section{Introduction}
\label{sec:intro}
% !TEX root = main.tex

Technical Debt (TD) has been defined by Cunningham as ``not-quite-right code'' \cite{Cunningham93}. Essentially, the terms refer to the debt an organization or an individual developing and releasing software should repay to make it acceptable, for example in terms of functionality, reliability, or maintainability.
Oftentimes, developers achieve awareness of the TD in their program, admitting it through comments, commit messages, or issues. This has been referred to as ``Self-Admitted Technical Debt'' (SATD) \cite{PotdarS14}.
Previous research has investigated why developers annotate software as SATD, essentially to keep track of what needs to be improved \cite{ZampettiFSP21}. Also, the analysis of SATD in existing programs has shown how developers take it seriously, as it gets removed in the majority of the cases \cite{MaldonadoASS17}, even though this often happens when the source code is completely replaced or even removed \cite{ZampettiSP08}.
\rev{As previous work found \cite{Bavota2016MSR,FucciCZNSP21,PotdarS14,ZampettiFSP21}, SATD relates to different problems in the program, such as the need to fix bugs occurring in 
certain circumstances, enhancing or even completing a feature, or improving the source code maintainability and quality in general.}

Researchers have proposed various kinds of approaches to aid developers with the management of SATD. On the one hand, while SATD comments are usually recognizable by commonly used keywords such as ``TODO'' or ``FIXME'', this is not always the case. Therefore, approaches leveraging several types of techniques ranging from simple regular expression matching \cite{PotdarS14} to shallow machine learning \cite{MaldonadoST17}, and deep learning \cite{RenXXLWG19} have been used to identify SATD comments. 

Also, some approaches recommend developers with the type of change that needs to be carried out to address the SATD~\cite{ZampettiSP20}.
While previous work has proposed approaches to guide developers towards paying back TD, to the best of our knowledge there is no specific work aimed at automatically resolving (SA)TD. Indeed, this can be a direction worthwhile to investigate, considering the significant advances in the application of generative approaches to software-related tasks, such as code completion \cite{ciniselli2021MSR,ciniselli2021empirical,svyatkovskiy2020fast,svyatkovskiy2020intellicode,li2017code}, program repair \cite{mashhadi-codebert,jiang2021cure,lutellier2020coconut,li2020repair}, vulnerability patching \cite{chen2022neural,fu2022vulrepair}, or code review \cite{tufano2021towards,tufano2022using,li2022codereviewer,li2022automating}. 
\rev{However, given the diversity of SATD, its repayment would require approaches that go beyond what has been already devised for each of the aforementioned tasks.}
Therefore, this paper aims to answer the following question:

\begin{quote}
	\emph{Can AI-based approaches automatically repay the technical debt?}
\end{quote}

To answer this question, we investigate the extent to which neural generative approaches based on deep learning transformers \cite{vaswani2017attention} can be used to repay SATD. 
An obvious question that can arise is whether SATD resolution is, in the end, equivalent to program repair. We believe there are a series of differences and challenges:

\begin{compactenum}
	\item As also pointed out by previous work \cite{ZampettiFSP21}, repaying (SA)TD not only means fixing bugs, but it also requires to improve the code in different ways, for example enhancing a feature, making the code able to handle certain special scenarios, or adopting alternative APIs. 
	\item Differently from other ``buggy'' code, SATD admits the presence of a problem, therefore directly highlighting the portions of the source code that need to be repaired.
	\item Neural models require a conspicuous amount of training data, which may be available for certain tasks (\eg code completion), and less for others, including SATD resolution. 
\end{compactenum}

To study how SATD can be addressed by generative models, we first created a dataset
of 5,039 SATD removal instances from 595 open-source projects by leveraging an available tool that detects SATD removals \cite{alomar2022satdbailiff}. Then, we leverage a pre-trained transformer model (CodeT5 \cite{wang2021codet5}) that previous work showed particularly effective  to cope with problems where the size of the training set is limited \cite{fu2022vulrepair,wang2022no}. Through different experiments, we test the effectiveness of several pre-training/fine-tuning strategies, including: 

\begin{compactenum}
	\item No pre-training, fine-tuning using SATD removal instances.
	
	\item Self-supervised pre-training followed by fine-tuning on SATD removals. Self-supervised pre-training exploits training objectives not requiring a supervised dataset. We use the \emph{masked language model} objective \cite{devlin2018bert,liu2019roberta}, which consists in providing the model with input sentences (\eg an English sentence, a \java method,  depending on the language of interest) having 15\% of their tokens masked, asking the model to predict them.
	
	\item Self-supervised and supervised pre-training followed by fine-tuning on SATD removal instances. Supervised pre-training can be used to pre-train the model on a task similar to the downstream one. In our case, we pre-train the model for the implementation of generic code changes, before fine-tuning it with SATD removals.
\end{compactenum}

%Once identified the best-performing combination of training objectives, 
We also experimented with the impact on the model's performance
%of providing it the SATD to fix 
with and without the SATD comment. 
This is worthwhile to study because (i) 
%The latter is worthwhile to experiment with for two reasons.
the SATD comment may act as a sort of prompt-tuning for the transformer \cite{vaswani2017attention} which has been shown to help models pre-trained on English text (such as CodeT5); and (ii) %to improve their performance. Second, 
a boost in performance motivates the usefulness of SATD comments not only as a trace for developers \cite{ZampettiFSP21}, but also as a way to aid AI-based approaches.
Finally, we experiment with the extent to which SATD can be addressed by leveraging a Large Language Model (LLM) chat bot, \ie ChatGPT \cite{chatgpt}. 

Results of our study indicate that automatically addressing SATD instances is a challenging task, and the best model we experimented with is able to correctly address 2.30\% (one attempt) to 8.10\% (ten attempts) of the SATD instances in our test set. Without any pre-training, the model is not able to address any SATD instance, likely due to the limited size of the fine-tuning dataset. Self-supervised pre-training helps in improving performance, which is further increased when the model is also subject to supervised pre-trained on a task (\ie implementing generic code changes) resembling the downstream one (\ie addressing SATD). 
%This is the best model we experimented with. 
Finally, we experimented with three different prompts for ChatGPT, with the best one being able to address only 1.19\% of the SATD in our dataset, confirming how challenging the tackled task is.
In summary, while the studied approaches can in some cases automatically repay SATD, there is still a long way to go to fully address this problem.

Overall, the paper contributes to the state-of-the-art on (SA)TD management and resolution with:
\begin{compactenum}
	\item An experimentation featuring seven different combinations of treatments on the use of pre-trained neural transformers for SATD repayment;
	\item Results of a study on the feasibility of using a LLM chat bot (ChatGPT) for SATD repayment; and
	\item A replication dataset that can be also used for further experiments in this area \cite{replication}.
\end{compactenum}

%The paper is organized as follows. \secref{sec:related} discusses the related literature.
%\secref{sec:study} details the design of our study, also providing a description of the approach used to automatically resolve SATD. Results are presented and discussed in \secref{sec:result}. Then, \secref{sec:threats} discusses the study's threats to validity, while \secref{sec:conclusion} concludes the paper and outlines directions for future work. Data availability information is in \secref{sec:replication}.

\section{Related Work}
\label{sec:related}
% !TEX root = main.tex

Given the emphasis of our investigation, we will focus our discussion on two primary research areas: (i) methods that assist in managing and removing SATD comments, and (ii) neural models that automate program repair \rev{and code review} tasks.
For a comprehensive overview of DL-based models applied in software engineering, we point the reader to the systematic literature review by Watson \etal \cite{watson2022systematic}.

\subsection{SATD Management and its Removal}
\label{sub:related-satd}

Previous works have studied SATD along various dimensions, and in particular, the types of SATD occurring in programs and its relation with software quality \cite{Bavota2016MSR}, the polarity of developers' comments with relationship to different SATD types \cite{CasseeZNSP22}, or the extent to which source code not containing an SATD may indeed require it \cite{ZampettiNAKP17}.
We only discuss work specifically related to SATD removal.  Further research about SATD can be found in a survey by Sierra \etal \cite{Sierra:jss2019}. 

SATD removal represents an important activity for software developers. Maldonado \etal \cite{MaldonadoASS17} analyzed the SATD removal in five Java open-source projects. Their findings indicate that the majority of SATD is being removed, about half of which by the same person introducing it.  A follow-up study by Zampetti \etal \cite{ZampettiSP08} performed a fine-grained analysis of SATD removals, finding that a large percentage of SATD removals occur ``accidentally'' along with other changes. Most of the changes required to remove SATD are complex ones, \ie requiring the addition, removal, or replacement of multiple source code lines. In some circumstances, SATD is removed by means of specific changes, \eg to conditionals in control-flow statements or to APIs.
\rev{In general, results of Zampetti \etal indicate that, beyond the cases where the code simply disappears, the type and span of changes required to address a SATD vary a lot, requiring in some cases simple code edits (\eg change a condition), and in other cases the addition of new blocks of code (\eg add new behavior), with the changes sometimes spanning in multiple places. This, unavoidably, makes the automated SATD fix a very challenging task.}

Despite previous work suggested that SATD is often removed ``by chance'', Tan \etal \cite{TanFA21} surveyed developers to understand whether they intentionally fix SATD. Their results indicate that in the majority of the cases, developers are conscious of the SATD removal and balance the pros and cons of this action. On the same line, Pina \etal \cite{PinaSG22} studied, through a survey, how developers prioritize technical debt management finding that, when developers decide to replay technical debt, they do it at their earliest convenience. Overall, the aforementioned work motivates ours.
% in different ways, by highlighting the importance of SATD removal for developers, and by pointing out, at least in some cases, certain regularities in SATD removal actions. 

The first work aimed at suggesting how to address SATD was proposed by Zampetti \etal \cite{ZampettiSP20}. They leveraged a model taking as input the SATD comment and the affected source code, to determine the type of change to apply---along six categories---for repaying the SATD.
%Convolutional Neural Network (CNN) working on the SATD comment and a Recurrent Neural Network (RNN) working on the affected source code. Their network classified the type of changes to be applied to remove the SATD into six categories. 
%The proposed approach achieved an overall Precision of 39\% and recall of 41\%. 
Differently from Zampetti \etal \cite{ZampettiSP20}, we propose to automatically recommend the actual fix instead of suggesting its category. This is made possible by the use of pre-trained models.

\subsection{\rev{Neural  Models for Automated Program Repair and Review}}

Researchers have leveraged various types of neural models for Automated Program Repair (APR). Chen \etal \cite{chen2019sequencer} proposed SequenceR, which is a vanilla version of Transformer with copy mechanism to handle the out-of-vocabulary problem (OOV), correctly predicting 950/4,711 ($\sim$20\%) fixes for the buggy component provided as input to the model. 
Tufano \etal \cite{Tufano:tosem2019} investigated the usage of Neural Machine Translation (NMT) to generate patches for buggy code, specifically \java methods. 
Their approach was trained using buggy methods provided as input, which were ``translated'' into their respective fixed versions.
%by the model through the necessary code changes. 
%The achieved results suggested the viability of NMT techniques for fixing bugs but also revealed limitations in handling complex methods with a high number ($>50$) of tokens. 
In a separate work, Tufano \etal \cite{tufano2019learning} leveraged a different NMT model to automatically apply code changes implemented by developers during pull requests (PRs). 
%Their investigation revealed that the proposed approach can replicate developer changes in PRs with a success rate of up to 36\%, but only when applied within a narrow context, \ie methods containing up to 100 \java tokens.

Jiang \etal \cite{jiang2021cure} suggested CURE, a GPT (Generative-Pretrained-Transformer) architecture \cite{radford2018improving} pre-trained on source code and using a subword tokenization technique to generate a smaller search space containing more correct fixes.

Mashhadi and Hemmati \cite{mashhadi-codebert} proposed the usage of CodeBERT \cite{feng2020codebert} to fix simple bugs in the context of \java programs. Once the model has been fine-tuned on the ManySStuBs4J dataset \cite{karampatsis2020often}, Mashhadi and Hemmati assessed its capabilities in generating meaningful fixes, finding that the devised approach can produce fixes for different types of bugs as a real developer would do in up to 72\% of cases. 

Mastropaolo \etal \cite{mastropaolo2021studying, mastropaolo2022tse} explored the capability of the Text-to-Text Transfer Transformer (T5) model in supporting various code-related tasks, including bug-fixing. %Their findings demonstrated the superior performance of T5 in comparison to non-pre-trained deep learning models \cite{Tufano:tosem2019} when producing fixes for buggy code components.

Lutellier \etal \cite{lutellier2020coconut} introduced CoCoNuT, a CNN-driven NMT model employing ensemble learning for producing bug fixes. The assessment conducted on four languages (Java, C, Python, and JavaScript) revealed that CoCoNuT successfully resolved 509 bugs, including 309 bugs not previously addressed by existing methods.

Li \etal \cite{li2020repair} proposed a tree-based RNN architecture to generate bug fixes. The usage of tree-based RNN allows modeling code and at the same time to learn tree-based structural code information from past bug fixes. 

\rev{Concerning the automation of code review activities on code, Tufano \etal \cite{tufano2021towards} made a first step towards automating code review tasks by utilizing Transformer models to implement code changes as requested by reviewers. In a subsequent study, Tufano \etal \cite{tufano2022using} further explored the use of pre-trained models and BPE-like schema tokenization \cite{kudo2018sentencepiece},  to advance the state-of-the-art of code review automation. Thongtanunam \etal \cite{thongtanunam2022autotransform} also confirmed the advantages of Transformer-based models and BPE-tokenization for improving the performance of DL-based systems in automating code reviews.	Other research in this domain includes CodeReviewer \cite{li2022codereviewer}, which utilizes a Transformer encoder-decoder model pre-trained with four tailored tasks for code review.}

Given existing \rev{APR and code review research} we empirically investigate how state-of-the-art approaches can be adapted to repay (SA)TD.

\rev{Different authors have proposed transformer models specifically adapted to support code edit tasks. Some of these works proposed AST-specific representations able to learn and predict code edits \cite{Brody0Y20,ChenHLMMTM21,YaoXYSN21}. However, these require the models to be trained from scratch, and to rely on an AST-based representation. 		
	A more general approach has been proposed by Zheng \etal \cite{ZhangP0LG22}, who introduced CoditT5. CoditT5 is based on the same architecture as CodeT5, and, given a sequence of tokens to repair, it predicts an edit plan, in terms of insert, delete, and replace operations. CoditT5 has been pre-trained on both natural language and source code and, with proper fine-tuning, showed superiority to other models (including CodeT5) for tasks such as comment updating, bug fixing, and automated code review.}

Recently, LLMs such as GPT-3 \cite{brown2020language} or GPT-4 \cite{openai2023gpt4} have propelled the automation of code-related tasks to new heights. Prenner and Robbes \cite{prenner2021automatic} investigated the extent to which the OpenAI's Codex Model \cite{chen2021evaluating} is able to localize and fix bugs.  The study's results highlighted the impressive performance of LLMs in zero-shot settings, exhibiting competitive results compared to the latest state-of-the-art techniques.

We take inspiration from the latest research and explore the capability of the new ChatGPT assistant \cite{openai2023gpt4} introduced by OpenAI. Specifically, we conducted experiments within a closed-setting scenario (\ie zero-shot learning) to evaluate the model's ability to generate code changes required when addressing \satd.

\section{Study Definition, Design and Planning}
\label{sec:study}
% !TEX root = main.tex
\label{sec:approach}

The \emph{goal} of this study is to evaluate DL-based solutions in automatically implementing code changes required to address SATD in \java code. The \emph{context} of the study features two deep learning models, namely CodeT5 \cite{wang2021codet5} and ChatGPT \cite{chatgpt}, and two datasets used for pre-training and fine-tuning the experimented models. The pre-training dataset features generic code changes implemented by developers in open-source projects and has been presented in the work by Tufano \etal \cite{tufano2019learning}. The fine-tuning dataset is a contribution of this paper and features SATD removal changes.

In the following, we formulate the study research questions (RQs) (\secref{sub:rqs}). Then, \secref{sub:datasets} describes the datasets used to train and test the experimented techniques. The latter are presented in \secref{sub:techniques}. We conclude by outlining the data analysis procedure in \secref{sub:data-collection}.

\subsection{Research Questions}\label{sub:rqs}

Given our overall goal (\ie assessing the capabilities of DL-based solutions in automatically addressing SATD), we formulate the following RQs:\smallskip

\textbf{RQ$_{1}$:} \textit{To what extent do pre-trained models of code support automated SATD repayment?} 
In RQ${1}$ we fine-tune the pre-trained CodeT5 \cite{wang2021codet5} model for the task of SATD repayment and assess its performance. We also investigate the role played by the self-supervised pre-training on CodeT5, \ie the extent to which the self-supervised pre-training helps in fixing SATD. RQ${1}$ provides a starting point for our investigation, showing what performance can be achieved by just fine-tuning an existing pre-trained model for SATD repayment.\smallskip

\textbf{RQ$_{2}$:} \textit{To what extent does the infusion of ``similar-task knowledge'' in pre-trained models of code benefits the automated SATD repayment?} 
While CodeT5 \cite{wang2021codet5} has been pre-trained using the \emph{masked language model} self-supervised objective, other forms of pre-training are possible. RQ$_{2}$ evaluates the effectiveness of performing a further pre-training step aimed at instilling in the model knowledge about a task resembling the downstream one (\ie the SATD repayment). This means that, before fine-tuning the model for SATD repayment, we leverage a supervised pre-training in which the model learns how to implement generic code changes. The rationale is that this task, while different from SATD repayment, can start driving the model's weights toward a configuration closer to the one needed for the downstream task.\smallskip

\textbf{RQ$_{3}$:} \textit{To what extent does the presence of ``context-specific knowledge'' help pre-trained models of code in the automated SATD repayment?} 

SATD instances can be represented as pairs $\langle comment, code\rangle$ where the $comment$ describes the SATD to address in the $code$. When assessing the performance of DL-based solutions in automatically addressing SATD, the $\langle comment, code\rangle$ pair represents the input of the model which is expected to produce a $revised\_code$ addressing the technical debt. If we factor out the $comment$ from the input the task becomes similar to those previously tackled in the literature through DL models, such as learning generic code changes implemented by developers \cite{tufano2019learning} or fixing bugs \cite{Tufano:tosem2019,jiang2021cure,mashhadi-codebert,lutellier2020coconut,chen2022neural}. For these tasks the model's input is just a $code$ in which a change (\eg a bug fix) must be implemented, thus producing as output the $revised\_code$. RQ$_{3}$ assesses the extent to which providing the SATD $comment$ to the model helps to address the TD, thus investigating whether training a SATD-specialized model is worthwhile as compared to just using a model trained to address generic code changes without relying on the SATD comment (see \eg \cite{tufano2019learning}).\smallskip

%\textbf{RQ$_{4}$:} \textit{Are domain-specific pre-trained language models of code zero-shot learners for SATD removal?}
%n RQ${4}$ we study the extent to which the infusion of domain-specific knowledge into CodeT5 unlock the possibility of using such a model in a zero-shot setting for the automatic removal of SATD comments.
%whether CodeT5 once infused with domain-specific knowledge (\ie how to address a generic code change) can be used without further fine-tuning to support the automatic removal of SATD. 
%A positive answer to RQ${4}$ would indicate that research on specialized models for SATD removal is unlikely to be relevant/beneficial.
%\smallskip

\textbf{RQ$_{4}$:} \textit{Are general-purpose large language models zero-shot learners for SATD repayment?}
In RQ${4}$ we study whether LLMs (and, in the specific case, ChatGPT \cite{chatgpt}) can be considered as out-of-the-box solutions for the automated  SATD repayment. While, for what concerns source code ChatGPT has been ``seen'' the entire GitHub, it has not been fine-tuned for the specific problem of SATD repayment. 
A positive answer to RQ${4}$ would indicate that research on specialized models for SATD repayment is unlikely to be relevant/beneficial.

\subsection{Context: Datasets}\label{sub:datasets}
We describe the pre-training and fine-tuning datasets used in our research, which are summarized in \tabref{tab:datasets}.

\subsubsection{Self-supervised pre-training on bi-modal data}
\label{sub:self-pretraining}
In the first three RQs, we employ CodeT5 \cite{wang2021codet5} as representative of a state-of-the-art pre-trained model of code. CodeT5 is a Text-To-Text Transfer Transformer (T5) model \cite{raffel2019exploring} pre-trained on code and natural language (\ie code comments).  \rev{Among different transformer models we have chosen CodeT5, as it has been used successfully for several tasks, including code summarization \cite{wang2021codet5}, source code generation \cite{zhou2022docprompting}, vulnerability patching \cite{fu2022vulrepair}, code review automation \cite{li2022automating}, code-to-code translation \cite{kusum2022unsupervised},  and shown to outperform other models such as CodeBERT \cite{feng2020codebert}, PLBART \cite{ahmad2021unified} and GraphCodeBERT \cite{guo2020graphcodebert}.}
%CodeT5 has been extensively employed in the software engineering literature \cite{wang2022no,chakraborty2022natgen,bui2022detect,troshin2022probing,zhou2023generation,le2022coderl,ahmed2022learning,wang2022compilable}. 

Wang \etal \cite{wang2021codet5} utilized the CodeSearchNet dataset \cite{husain2019codesearchnet} for pre-training CodeT5 using the masked language model objective (\ie self-supervised pre-training by randomly masking 15\% of the input asking the model to predict it). This dataset includes functions written in six different programming languages (Go, Java, JavaScript, PHP, Python, and Ruby). A subset of these functions also includes a top-level comment (\eg Javadoc for Java). In addition, Wang \etal gathered extra data from C/C\# repositories hosted on GitHub. This led to a total of 8,347,634 pre-training code functions: 3,158,313 of these functions are paired with their documentation, while 5,189,321 consist solely of code.

\subsubsection{Supervised pre-training on generic code changes}
\label{sub:ft-dataset-code-changes}

Tufano \etal \cite{tufano2019learning} proposed the usage of NMT to learn how to automatically apply code changes implemented by developers during pull requests (PRs). The NMT model has been trained on a dataset featuring PRs from three \emph{Gerrit} \cite{gerrit} repositories: (i) \emph{oVirt}, (ii) \emph{Android}, and (iii) \emph{Google}. 

Each of these repositories groups multiple projects (\eg all those related to the \emph{Android} operating system), with the authors focusing on the ones written in \java. For each PR, Tufano \etal extracted two versions of the files involved in the change diff: The version before the PR was implemented and the version after the PR has been merged. These files have then been parsed to extract a total of 631,307 pairs of methods $\langle m_b, m_a\rangle$ representing the same method before ($m_b$) and after ($m_a$) the changes implemented in the PR. The idea was to train the NMT model on these pairs to see if it was able to learn generic code changes that developers might implement in the context of PRs. 

We leverage this dataset in the context of RQ$_2$ to investigate whether the infusion of ``similar-task knowledge'' in pre-trained models of code benefits the automated SATD repayment. Starting from the $\sim$630k pairs in the original dataset we discarded instances containing non-ASCII tokens, and those having $\#tokens$ > 1024. The latter filter is necessary to manage the computational complexity of training large DL-models and is a common practice in the software engineering literature \cite{mastropaolo2022using,ciniselli2021empirical,tufano2022using,tufano:testGeneration,tufano2021towards,tufano2019learning}. For example, in the original work by Tufano \etal \cite{tufano2019learning} in which this dataset has been created, the authors removed all pairs having $\#tokens$ > 100. Subsequently, we eliminate duplicated pairs $\langle m_b, m_a\rangle$, obtaining a final dataset of 284,190 instances. We split the processed set of methods into 90\% training and 10\% validation. The former will be used to perform supervised pre-training on the task of learning generic code changes. The latter is instead employed to identify the best-performing checkpoint while performing the supervised pre-training.

\begin{table}[h!]
	\centering
	\caption{Number of instances included in the datasets we used to train, test and evaluate the models}
		\label{tab:datasets}
		\begin{adjustbox}{max width=\columnwidth}
		\begin{tabular}{lrrrr}
			\hline
			\vspace{0.03cm}
			\textbf{Dataset}  & \textbf{Train} & \textbf{Test} & \textbf{Eval}  & \textbf{Overall}\\
			\hline
			
			Generic code changes \cite{tufano2019learning} & 255,771 & - & 28,419 & 284,190\\
			\midrule
			SATD removal  & 3,537 & 1,000 & 502 & 5,039 \\
			\midrule
		   Total & 259,308 & 1,000 & 28,921 & 289,229 \\
			\bottomrule
		\end{tabular}
	\end{adjustbox}
\scriptsize

\end{table}

\subsubsection{Fine-tuning dataset of SATD removals}
\label{sub:ft-dataset}

As a first step to build the fine-tuning dataset, we collected a list of \java repositories leveraging the GitHub Search tool by Dabic \etal \cite{dabic2021sampling}. The querying user interface allows to identify GitHub projects that meet specific selection criteria. We selected all \java projects having at least 500 commits, 10 contributors, 10 stars, and not being forks (to reduce the chance of mining duplicated code). The commits/contributors/stars filters aim at discarding personal/toy projects. Instead, the decision of narrowing down the scope to only \java as a programming language was dictated by (i) the will of reusing the previously described \java dataset by Tufano \etal (\secref{sub:ft-dataset-code-changes}) for supervised pre-training, and (ii) the usage in our toolchain of tools only supporting Java (\eg \tool, as described in the following). We collected 6,971 \java repositories.

To extract changes aimed at addressing SATD instances we rely on the \tool tool by AlOmar \etal \cite{alomar2022satdbailiff}. \tool can identify commits featuring the addition, removal, or change of SATD instances in the history of \java git repositories. We are interested in mining from the history of the subject \java repositories commits featuring removal events, since those are the ones implementing changes aimed at addressing SATD and, thus, are the ones suitable for our fine-tuning. Unfortunately, we found this extraction process to be extremely expensive. For this reason, we set two boundaries for our data mining. First, we allowed \tool to process each repository for at most 60 minutes. Indeed, we observed that for some very large projects the mining procedure could go on for days. If a repository could not be analyzed within 60 minutes, the repository was discarded from our study. Second, we set 50 days as the maximum boundary for the mining procedure. In this period, \tool successfully analyzed the entire history of 809 \java projects hosted on GitHub.

These 809 \java projects yielded a total of 519,440 SATD-related events including \emph{SATD\_REMOVED}, \emph{SATD\_ADDED}, and \emph{SATD\_CHANGED}. We extracted the 139,803 concerning \emph{SATD\_REMOVED}.
Then, we further refined this list by only selecting instances for which the SATD comment extracted by \tool contained specific keywords describing the presence of \satd: \texttt{to(-)do},  \texttt{fix(-)me}, \texttt{check(-)me}, \texttt{hack(-)me}, and \texttt{xxx}. In principle, we are aware that such further filtering may reduce the dataset construction recall and, ultimately, the dataset size. However, as the SATD detection performed by \tool is based on a ML-based approach, it is inherently subject to false positives, and we wanted to avoid experimenting with changes that were not related to SATD removal.
We have chosen the aforementioned keywords since those are well-known patterns used to signal SATD \cite{PotdarS14,CasseeZNSP22}.
After this further pruning, we obtain a dataset where each instance is a triplet \texttt{commit\_before}, \texttt{commit\_after}, and \texttt{comment}, where the two commits indicate the version of the code affected (\texttt{commit\_before}) by and cleaned from (\texttt{commit\_after}) the SATD, while \texttt{comment} is the code comment documenting the SATD which is linked to a specific \java file.

We removed duplicated instances, namely those being characterized by the same triplet (\eg due to the same SATD fixed in the same commit in different files). This left us with 75,083 instances which could feature the SATD in any part of the impacted \java file. However, we are interested in training the DL-model to address SATDs affecting a specific method, ignoring SATD instances related to \eg class instance variables, {\tt import} statements \etc The reason for such a choice is two-fold. First, also the pre-training datasets are defined at function-level granularity, thus suggesting a similar fine-tuning to take full advantage of the knowledge acquired during pre-training. Second, providing an entire \java file as input to the model makes the training extremely expensive, since the length of the input sequences will grow to tens of thousands of tokens. Thus, we parsed the code file in the \texttt{commit\_before} to see if the SATD comment was within a method $m_{satd}$ or immediately above it. This was the case for 65,380 instances. 

\begin{figure}[t]
	\centering
	\includegraphics[width=0.9\columnwidth]{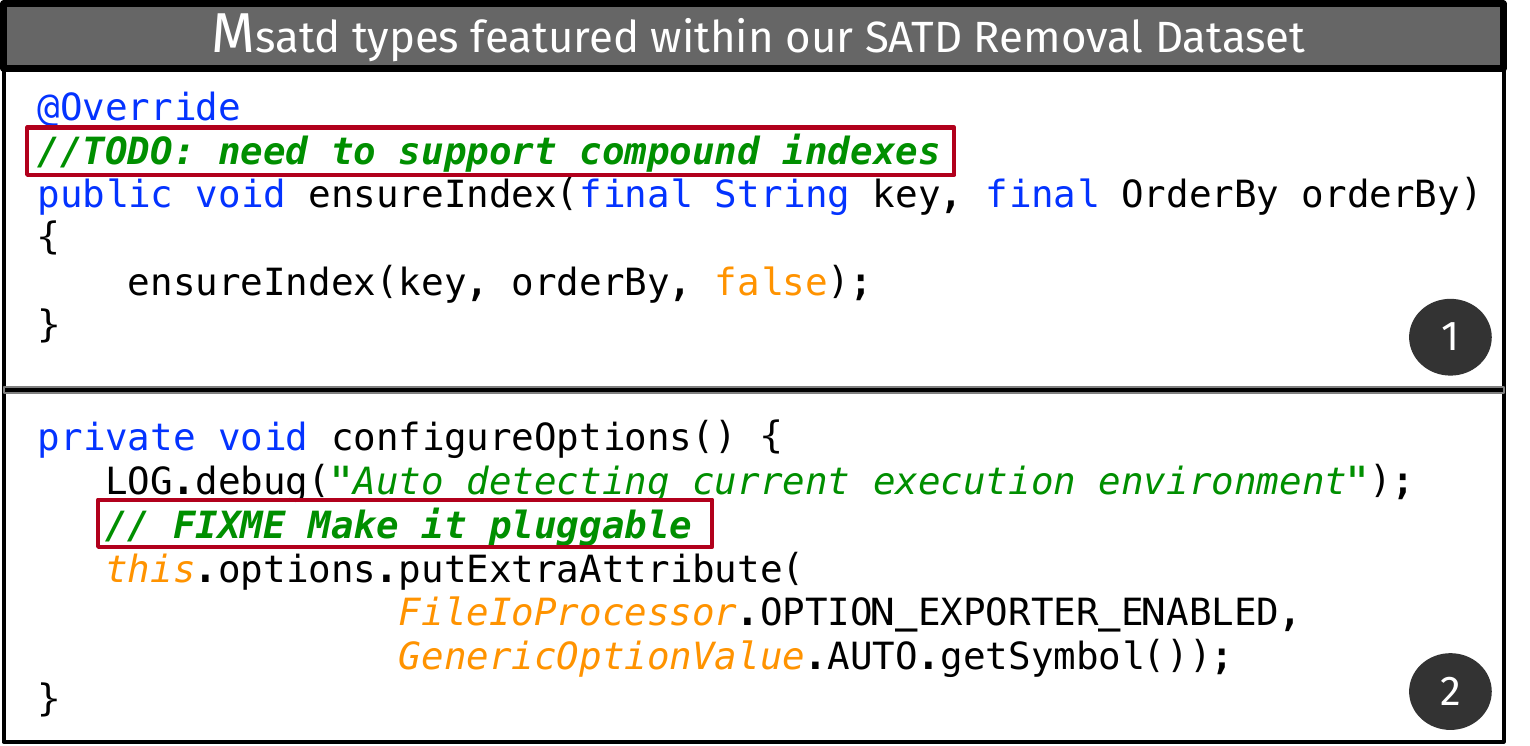}
	\caption{SATD removal instances in our fine-tuning dataset}
	\label{fig:example-dataset-supported}
	\vspace{-0.5cm}
\end{figure}

Scenario \circled{1} depicted in \figref{fig:example-dataset-supported} shows a SATD comment preceding the implementation of \texttt{ensure\-Index} method, while in \circled{2} the comment documenting the TD is included within the body of the \texttt{configure\-Options} method.
\rev{Based on what we have explained so far, we work on the following assumption: 
	\begin{quote}
		If an SATD comment is removed, the changes performed within the same commit and in the method to which the SATD comment was attached are related to repaying such an SATD.
	\end{quote}	
}

Given the available dataset, our aim is to create the final training triplets $\langle m_{satd}, m_{fixed}, comment\rangle$, where $\langle m_{satd}, comment\rangle$ represents the model's input and $m_{fixed}$ the model's output. To achieve this goal, some additional checks are required. First, as previous work has found \cite{ZampettiSP08}, it is possible that addressing the SATD requires the deletion of $m_{satd}$, or the implementation of other methods while leaving $m_{satd}$ unchanged (except for the removal of the SATD comment). Thus, we verify that (i) a method having the exact same name of $m_{satd}$ exists in \texttt{commit\_after}, and (ii) by removing the SATD comment from $m_{satd}$ we obtain a method $m_{satd}' \neq m_{fixed}$. The first filter guarantees that $m_{satd}$ still exists in \texttt{commit\_after}, while the second ensures that changes have been implemented in $m_{satd}$ to address the SATD (obtaining $m_{fixed}$). This cleaning left us with 12,267 triplets. 

We manually inspected 100 triplets to look for additional problematic cases. We found triplets characterized by ``meaningless'' SATD comments, such as ``TODO'' not followed by anything else. These comments do not really describe a SATD and, for this reason, we decided to exclude from our dataset all triplets having a $comment$ featuring less than three words that are unlikely to describe a SATD in enough detail to be understood (8,564 instances left). 

\begin{figure}[t]
	\centering
	\includegraphics[width=0.9\columnwidth]{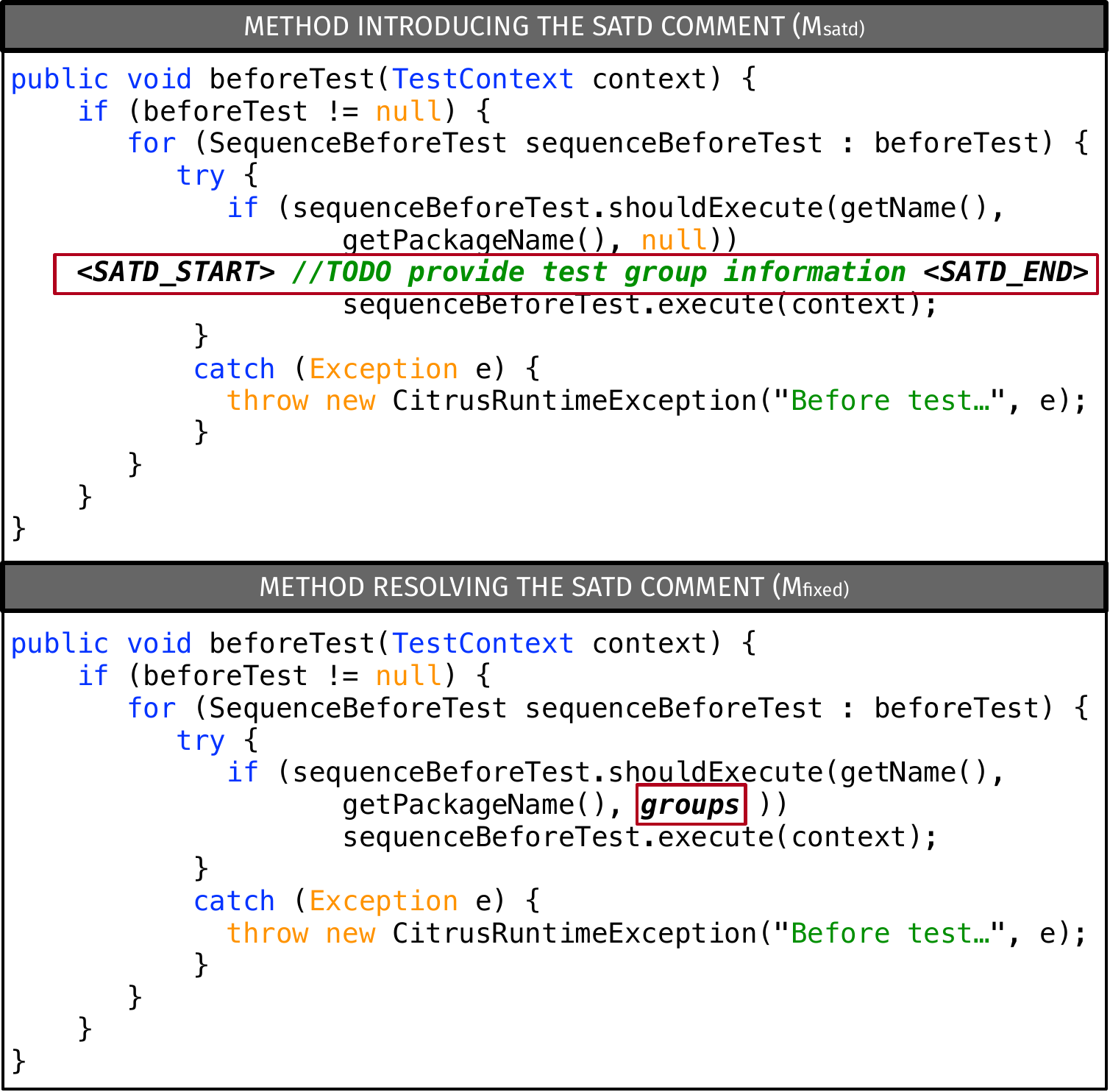}
	\caption{SATD removal instance in our fine-tuning dataset}
	\label{fig:example-dataset}
	\vspace{-0.5cm}
\end{figure}

Finally, we used the code-tokenize Python library \cite{richter2022tssb} to extract a tokenized version of the extracted methods and removed triplets featuring methods having $\#tokens$ > 1024, and instances that raised errors while being tokenized. After filtering, we ended up with a total of 5,039 triplets derived from 595 \java projects, which constitute our fine-tuning dataset. The latter is further split into 70\%, 20\%, and 10\% for training, testing, and validation of the models, respectively. These triplets have been processed to introduce two special tokens $\langle SATD\_START \rangle$ and $\langle SATD\_END\rangle$ which serve to tag the start and end of the SATD comment within $m_{satd}$. As previous work did \cite{fu2022vulrepair,tufano2022using}, the idea is that these tokens could help direct the model's attention toward relevant sections of the input. \figref{fig:example-dataset} depicts an example of instance from the dataset we built. 

Note that we create two different versions of the SATD removal dataset, both containing the same number of instances across training, testing, and validation. 

However, while the first one also contains the SATD comment, the latter is removed in the second one. This is necessary to address $RQ_{3}$, \ie to determine the extent to which admitting TD would not only serve as a trace for the developers but also as an aid for automated tools.

\subsection{Experimented Techniques} \label{sub:techniques}
As we are interested to study the performance of different DL-based solutions for automatically addressing SATD instances, we focus on two recently presented models, namely CodeT5 \cite{wang2021codet5} and ChatGPT \cite{chatgpt}. 
For the former, we use the $CodeT5_{base}$ variant, featuring 220 million trainable parameters. We use the default architecture and hyperparameters of $CodeT5_{base}$ featuring 12 Transformer Encoder blocks, 12 Transformer Decoder blocks, 768 hidden sizes, and 12 attention heads. The learning rate is set to 2e-5.

While CodeT5 has been specifically pre-trained and fine-tuned to support software engineering tasks, ChatGPT is a general-purpose LLM designed and developed by OpenAI to produce human-like responses for a broad spectrum of language-related tasks (\eg question-answering, language translation, coding tasks, \etc). There are currently two versions of ChatGPT, one built on top of GPT-3.5 and one exploiting GPT-4.0. Given the current restrictions on the usage of GPT-4.0, we carried out our research using GPT-3.5 as the foundational model for ChatGPT. Although considered ``less proficient'' than the chat-bot built using GPT-4.0, the version employed for our experiments, with 154 billion parameters model (GPT-3.5), is still in the LLM category, and it is definitely by far a larger model than CodeT5.

In the following, we detail how we used the two models to answer our RQs. In particular, we pre-trained, fine-tuned, and queried the models using several different strategies. All fine-tunings have been executed for a maximum of 50 epochs on the ``SATD removal'' dataset (see \tabref{tab:datasets}). To cope with overfitting, we stop the fine-tuning using an early stopping procedure assessing the loss of the model on the validation set every epoch, using a delta of 0.01 and patience of 5. This means that the training process stops if a gain lower than delta (0.01) is observed after 5 consecutive epochs and the best-performing checkpoint up to that training step is selected.

\subsubsection{No Pre-training + Fine-tuning (RQ$_1$)}
We fine-tune on the context-specific SATD removal dataset (\ie the one including the comment documenting the SATD) a T5$_{base}$ model \cite{raffel2019exploring} (\ie the same used for CodeT5 \cite{wang2021codet5}) without any pre-training. \rev{To this aim, we start by randomly initializing the weights of the model, which will be adjusted during the fine-tuning procedure.} Such a model serves as a baseline to assess the impact of different pre-trainings on the model's performance. 

\subsubsection{Self-supervised Pre-training + Fine-tuning (RQ$_1$)}
\label{sub:codet5-finetuning}
We start from the CodeT5 model pre-trained using a self-supervised objective (\ie \emph{masked language model}) and fine-tune it on the context-specific SATD removal dataset. 

\subsubsection{Self-supervised \& Supervised Pre-training + Fine-tuning (RQ$_2$ and RQ$_3$)}
\label{sub:code-changes-model}
Previous works in the natural language processing \cite{zhang2020pegasus} and in the software engineering literature \cite{chen2022neural,wei2019code} suggest that exploiting a supervised pre-training objective that resembles the downstream task (in our case, SATD repayment) can play a positive role on the models' performance. For this reason, we further pre-trained CodeT5 for five epochs using the ``generic code changes'' dataset described in \secref{sub:ft-dataset-code-changes}. 
Following that, we continue fine-tuning CodeT5, which has been enhanced with domain-specific knowledge, on the SATD removal dataset, both with and without additional context, \ie code comments.

\subsubsection{Zero-Shot Prompt Tuning (RQ$_4$)}
\label{sub:prompt-tuning}
We designed three prompt templates aimed at querying ChatGPT for the SATD repaying task. All prompts feature the SATD $comment$ and the method affected by SATD ($m_{satd}$):

\begin{enumerate}
	\item \texttt{Remove this SATD:  \{$comment$\} from the following code  \{$m_{satd}$\}} 
	
	\item \texttt{Perform removal of this SATD: \{$comment$\} from this code \{$m_{satd}$\}} 
	
	\item \texttt{This code \{$m_{satd}$\} contains the following SATD: \{$comment$\} remove it} 
	
\end{enumerate}

We also tried to explain ChatGPT the notion of SATD before querying it with any of the three above-listed prompts. However, we did not observe significant changes in the output, thus indicating that ChatGPT is ``aware'' of what SATD is.

\subsection{Data Collection and Analysis}
\label{sub:data-collection}
We run each trained CodeT5 on the 1,000 \java methods in the test set, asking it to implement the code changes needed to repay the SATD. We use the beam search decoding schema \cite{freitag2017beam} to produce multiple candidate repayments for an input $m_{satd}$. In the case of ChatGPT, we use the OpenAI APIs to query it. However, the APIs do only allow collecting a single answer (solution) from ChatGPT.

In the following, we summarize the seven different models' configurations we experiment with:

\begin{compactitem}
	\item 1: \emph{No pre-training + context-specific fine-tuning}, in the results referred as M0;
	
	\item 1: \emph{CodeT5 + context-specific fine-tuning}, referred as M1;
	
	%\item 1: \emph{CodeT5 + supervised pre-training on code changes (\ie zero-shot learning setting)}, referred as M2;
	
	\item 1: \emph{CodeT5 + supervised pre-training on code changes + context-specific fine-tuning}, referred as M2$_{CC}$;
	
	\item 1: \emph{CodeT5 + supervised pre-training on code changes + no-context fine-tuning}, referred as M3$_{CC-Ablation}$;
	
	\item 3: \emph{ChatGPT in zero-shot learning setting} $\times$ 3 prompt templates ($M4_{T1-T3}$), where the digit 1-3 indicates the used template among those described in \secref{sub:prompt-tuning}.
	
\end{compactitem}

We assess the performance of each model using two metrics. First, the percentage of Exact Match predictions for different beam sizes $K$ (EM@K), namely the cases in which the generated output is identical to the expected $m_{fixed}$. For CodeT5 we experiment with $K$ equal 1, 3, 5, and 10. For the reasons previously explained, we only computed EM@1 for ChatGPT. Second, we compute the CrystalBLEU score \cite{eghbali2022crystalbleu} between the generated predictions and the $m_{fixed}$ target. CrystalBLEU measures the similarity between a candidate (predicted code) and a reference code (oracle), similar to how the BLEU score \cite{papineni2002bleu} measures similarity between texts. However, CrystalBLEU is specifically designed for code evaluation, while retaining desirable properties of BLEU, specifically being language-agnostic and minimizing the effect of trivially shared $n$-grams, which would produce inflated results. 

To better understand the extent to which the considered techniques can successfully address SATD, we analyze the edit actions (\ie deleting, adding, moving, or changing) to code elements required in each SATD repayment instance. To this aim, we use the
Gumtree Spoon AST Diff \cite{FalleriMBMM14} to gather the \texttt{Delete}, \texttt{Insert}, \texttt{Move}, and \texttt{Update} actions performed on the source code AST nodes when SATD is being addressed. Specifically,  we compute the \emph{actual} AST edit actions, \ie those obtained by differencing the \texttt{input} and the \texttt{target} (\ie ground truth) of the model.

Subsequently, we create two separate buckets. The first bucket includes all methods where the best-performing model accurately addresses the SATD comment, while the second bucket includes methods for which the suggested code is inconsistent with the developer's proposed repayment (\ie ground truth). For both categories, we present the relative counts of AST edit actions learned by the model (when the code suggested for addressing the SATD is actually correct) and those where the model faces difficulties in providing a significant implementation. 

Also, we perform statistical tests to determine whether one of the experimented techniques is more effective in producing code changes to address SATD.  We use McNemar's test \cite{mcnemar} (with is a proportion test for dependent samples) and Odds Ratios (ORs) on the EMs that the techniques generate. We also statistically compare the distribution of the CrystalBLEU scores (computed at the sentence level) for the predictions generated by each technique by using the Wilcoxon signed-rank test \cite{wilcoxon}. The Cliff’s Delta (d) is used as effect size \cite{Cliff:2005} and it is considered: negligible for $|d|$  0.10, small for 0.10 $\le$ $|d|$ < 0.33, medium for 0.33 $\le$ $|d|$ < 0.474, and large for $|d|$ $\ge$ 0.474. For all tests, we assume a significance level of 95\% and we account for multiple tests by adjusting $p$-values using Holm’s correction procedure \cite{Holm1979a}.

Finally, we discuss examples of successfully addressed SATD comments by the top-performing model and, at the same time, we present cases where the model was unable to pay back the TD.

%AST Chages quantiative analysis!

%Qualitative --> just pictures from the quantitative evaluation

%We complement the quantitative analysis by manually analyzing \textcolor{red}{XXX} elements, \textcolor{red}{YYY} where the model accurately implemented the intended code changes and \textcolor{red}{YYY} where the implemented code changes do not align with the oracle's provided implementation (\ie ground truth).
%In addition to this analysis, we also shed light on the true potential of the models in suggesting removal for SATD comment by categorizing the SATD types using the detailed taxonomy presented by Cassee \etal \cite{cassee2022self}. To this end, we use the same set of \textcolor{red}{XXX} instances already employed to conduct the above detailed investigation. 
%Hence, we label \textcolor{red}{YYY} elements for which the model proposed the appropriate change for the removal of the SATD comment and \textcolor{red}{YYY} instances where the recommendation deviated from the changes implemented by the developer.

\section{Results}
\label{sec:result}

\begin{table*}[ht!]
	\caption{Exact Match (\ie the recommended code is equal to the oracle) and CrystalBleu scores achieved by the different techniques when addressing SATD comments. In bold we report the highest value for both metrics when producing $K$=1, $K$=3, $K$=5, and $K$=10 candidate removals. \vspace{-0.2cm}}
	\label{tab:perfect}
	\resizebox{\textwidth}{!}{%
		\begin{tabular}{lccccc>{\columncolor[gray]{0.8}}rr>{\columncolor[gray]{0.8}}rr>{\columncolor[gray]{0.8}}rr>{\columncolor[gray]{0.8}}rr}
			\hline
			\multirow{2}{*}{{\bf Model}} &  \multicolumn{4}{c}{{\bf Training Configuration}} & 
			&   \multicolumn{2}{N}{\bf Top-1} 
			&  \multicolumn{2}{N}{\bf Top-3}  
			&  \multicolumn{2}{N}{\bf Top-5} 
			&   \multicolumn{2}{N}{\bf Top-10} \\   \hhline{~----~--------}
			& {\bf Self-supervised PT} & {\bf Supervised PT} & {\bf SATD Comm.} & {\bf FT} &
			&  {\bf EM} & {\bf CB}  
			&  {\bf EM} &  {\bf CB}
			& {\bf EM} &  {\bf CB}
			& {\bf EM} & {\bf CB} \\
			\hline
			
			M0 &  \xmark & \xmark & \cmark	& \cmark & & 0\%   & 22.13\%               & 0\%   & 25.98\%            &  0.0\%   & 25.43\%       &   0.0\% &  25.14\%   \\
			
			\hline
		   M1 & \cmark & \xmark & \cmark & \cmark & &  2.23\%   & 73.11\%             & 5.40\%     & 73.95\%         & 6.10\%   & 74.19\%       & 7.20\%  & 73.87\%         \\

			\hline
		   M2$_{CC}$  &\cmark & \cmark & \cmark & \cmark & & \cellcolor[HTML]{656565}\color[HTML]{FFFFFF}  \bf 2.30\%   & \bf 73.41\%              &    \cellcolor[HTML]{656565}\color[HTML]{FFFFFF}  \bf 5.60\%   & \bf 74.20\%          &   \cellcolor[HTML]{656565}\color[HTML]{FFFFFF}  \bf 6.70\%  & \bf 74.28\%        & \cellcolor[HTML]{656565}\color[HTML]{FFFFFF}  \bf 8.10\%     & \bf 74.25\%       \\
		   
		   \hline
		   M3$_{CC-Ablation}$  &\cmark & \cmark & \xmark & \cmark & &  0.9\%   &  72.58\%              &   2.90\%   &  73.48\%          &   3.80\%  &  73.60\%         &   5.10\%     & 73.50\%       \\
			
			\hline
		   M4$_{T1}$ &\cmark & \xmark &\cmark & \xmark && 1.18\%   & 37.30\%              &  -   & -          &  -   & -      &     -   & -         \\
			
			\hline
			
			M4$_{T2}$ &\cmark & \xmark &\cmark & \xmark& & 1.19\%   & 49.68\%           &  -   & -          &  -   & -      &     -   & -         \\
			\hline
			
			M4$_{T3}$ & \cmark & \xmark &\cmark & \xmark && 0.0\%   & 1.70\%              &  -   & -          &  -   & -      &     -   & -         \\
			
			\hline
		
		\end{tabular}
	}
	\vspace{-0.3cm}
\end{table*}

\begin{table}[t]
\centering
\caption{Comparison among different techniques for top-1 predictions: McNemar's and Wilcoxon's test results.\vspace{-0.2cm}}
\label{tab:stats}
\begin{adjustbox}{max width=\columnwidth}
\begin{tabular}{lrr|rr}
  \hline
  \multirow{2}{*}{\textbf{Comparison}} &  \multicolumn{2}{c}{\textbf{McNemar's Test}} & \multicolumn{2}{c}{\textbf{Wilcoxon's Test}} \\ \cline{2-5}
 & \textbf{\emph{p}-value} & \textbf{OR}  & \textbf{\emph{p}-value} & \textbf{d} \\ 

\hline

M1 vs. M0 & - & - & $<$0.05 & -0.86 (L) \\ 
\hline 
M1  vs. M2$_{CC}$ & $>$0.05 & 1.0 & $>$0.05 & -0.01 (N) \\ 
\hline
 M3$_{CC-Ablation}$ vs. M2$_{CC}$  &  $<$0.05 & 8.0 & $<$0.05 & -0.01 (N) \\ 
\hline
M4$_{T1}$ vs.  M2  &  $>$0.05  & 1.38 &  $<$0.05  & -0.54 (L) \\ 
M4$_{T2}$ vs. M2  &  $>$0.05& 1.36 &  $<$0.05  & -0.43 (S) \\ 
M4$_{T3}$ vs. M2 &  - & - &  $<$0.05  & -0.98 (L) \\ 
  
  \hline
\end{tabular}
\end{adjustbox}
\vspace{-0.3cm}
\end{table}

\tabref{tab:perfect} reports the results obtained by the studied techniques when addressing SATD instances from our test set.  The first column (``Model'') provides a unique identifier we assigned to each of the 7 experimented techniques described in \secref{sub:data-collection}.
The ``Self-supervised PT'' and ``Supervised PT'' indicates whether a specific configuration we tested featured the two types of pre-training, where the self-supervised is the one adopting the \emph{masked language model} objective and the supervised uses the \emph{generic code changes} dataset to provide the model with knowledge about changing code. The ``SATD Comm.'' column indicates whether the fine-tuning (or the prompting in the case of ChatGPT) included the SATD comment in the model's input, while the ``FT'' column shows which model has been fine-tuned on our ``SATD removal'' dataset. Lastly, EM and CB indicate the performance of a specific configuration in terms of (i) the percentage of predictions that are Exact Matches (EM), and (ii) the average CrystalBLEU score (CB) \cite{eghbali2022crystalbleu} for all predictions in the test set. We present the results for different beam sizes ($K$) of 1, 3, 5, and 10.

\tabref{tab:stats} reports the results of the statistical tests (McNemar's test and Wilcoxon signed-rank test), with adjusted $p$-values, OR, and Cliff's $d$ effect size. An $OR>1$, or a positive Cliff's $d$ indicate that the right-side treatment outperforms the left-side one.  To enhance readability, when doing the comparisons, we arranged the treatments to display ORs $\geq 1$.

\subsection{RQ$_{1}$: To what extent do pre-trained models of code support the automated SATD repayment?} 

The first two rows of \tabref{tab:perfect} report the outcomes of the non-pre-trained model (M0) and its counterpart using self-supervised per-training on code and technical language (M1). 
While M0 is unable to address any SATD (EM = 0.0\%) for all values of $K$, the prediction performances of M1 range from 2.23\% ($K$=1) to 7.20\% ($K$=10). M1 is a CodeT5 fine-tuned for SATD removal and our results stress the importance (and validity) of the pre-training procedure performed on it by the original authors \cite{wang2021codet5}. Also, the difference in CrystalBLEU with respect to the non-pre-trained model (M0) is statistically significant ($p$-value < 0.05), according to  Wilcoxon signed-rank test, and is accompanied by a \emph{Large} Cliff's Delta. In the absence of EMs for M0, McNemar's test results cannot be computed.

\begin{resultbox}
	\textbf{Answer to RQ$_{1}$.} The use of a self-supervised pre-trained model (CodeT5) has a significantly positive benefit when addressing SATD, if compared to a non-pre-trained model. The latter is unable to produce exact matches, likely due to the limited size of the fine-tuning dataset.
	%which employs a self-supervised training objectives, such as masked language models has a significantly positive effect when it comes to automatically addressing SATD comments.
	%In contrast, techniques that do not depend on pre-training procedures (\eg M0) experience lower performances primarily because of the limited size of the  fine-tuning dataset which impairs the model's capacity to learn the language of interest (as evidenced by the CrystalBleu scores) and, at the same time adequately addressing SATD comments.
\end{resultbox}

\subsection{RQ$_{2}$: To what extent does the infusion of ``similar-task knowledge'' in pre-trained models of code benefits the automated SATD repayment?} 

Instilling task-similar (\ie code changes) knowledge into the model (M2$_{CC}$, featuring both self-supervised and supervised pre-training) results in a slight performance improvement as compared to M1 (\ie self-supervised pre-training only). While there is an improvement across all beam sizes (see \tabref{tab:perfect}), this is usually minor. For example, when only relying on the top prediction (\ie $K$=1), the EMs rise from 2.23\% to 2.30\% which, given the 1,000 instances featured in our test set, means 7 new EM predictions. The improvement is slightly higher when looking at higher values of $K$, with a +0.9\% reached for $K$=10 (7.20\% vs. 8.10\%). Upon statistically comparing both models (M1 vs. M2$_{CC}$), the McNemar's test (\tabref{tab:stats}) reports a lack of significant differences ($p$-value > 0.05) in EM predictions between M1 and M2$_{CC}$. Wilcoxon signed-rank test also suggests non-significant differences in the distributions of CrystalBLEU scores between M1 and M2$_{CC}$. Such a result is in line with what was observed by Tufano \etal \cite{Tufano:icse2023}. 
%in their empirical investigation on the impact of pre-training objectives on the performance of DL-based solutions for code-related tasks. 
They found that adopting a pre-training objective resembling the downstream task does not always substantially help, questioning the effort needed for the additional training time. This also seems to be the case when addressing SATD. Despite this, M2$_{CC}$ still is the best-performing model we experimented with and, for this reason, we performed some additional analyses on its predictions. 

\begin{figure*}[t]
	\centering
	\includegraphics[width=0.86\textwidth]{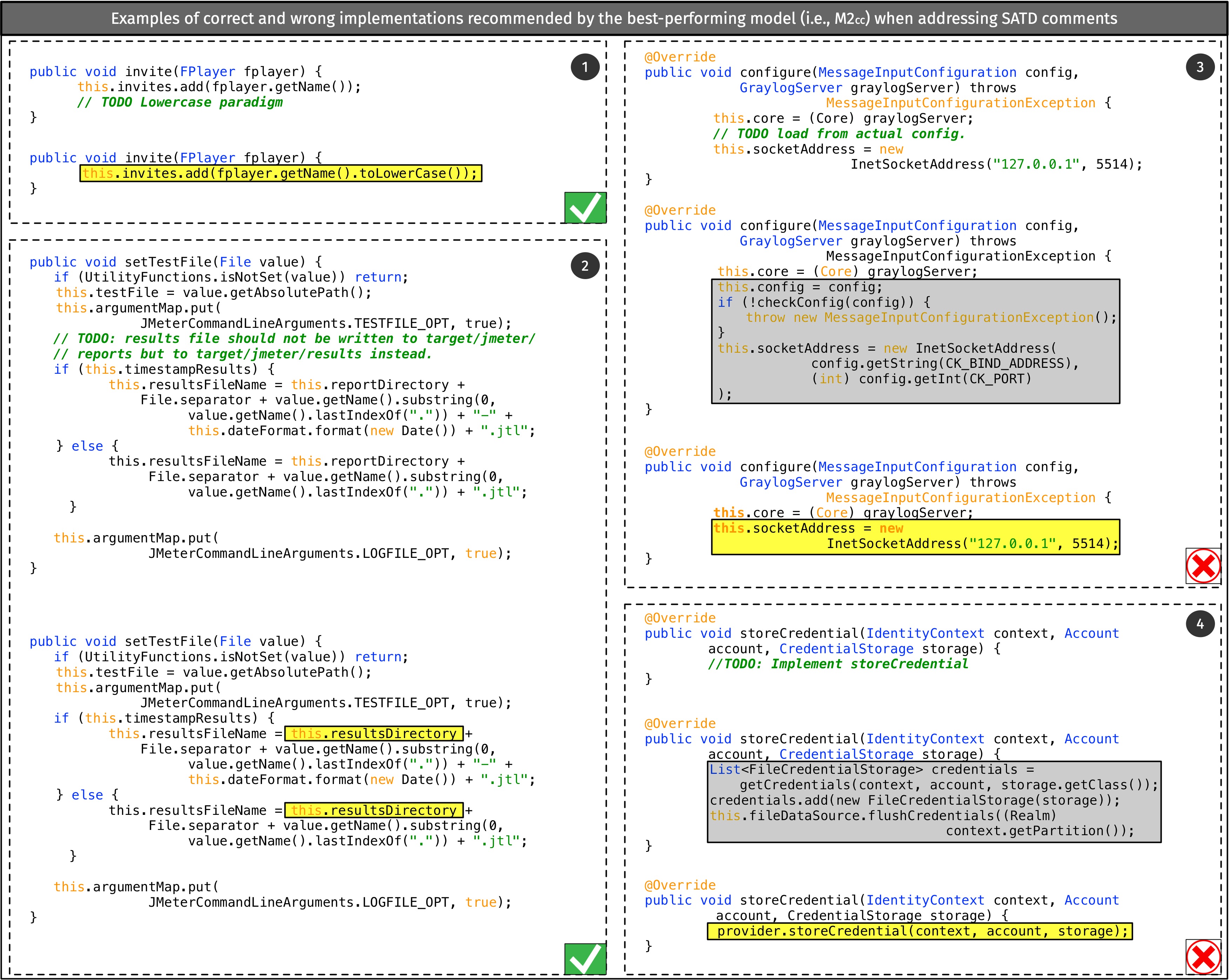}
	\caption{Example of 4 SATD comments addressed by the model (2 correct and 2 wrong) for top-10 candidate recommendations. }
	\label{fig:qualitative}
	\vspace{-0.4cm}
\end{figure*}

The EM predictions generated by M2$_{CC}$ with $K$=10, feature a total of 768 AST \emph{edit actions} correctly implemented by the model. Out of these, 62.36\% are \texttt{Delete} operations (\ie an AST node is removed), 33.60\% are \texttt{Inserts} (\ie a new node is introduced into the AST), and 2.34\% and 1.70\% are \texttt{Move} and an \texttt{Update} operations, respectively \includegraphics[scale=0.17]{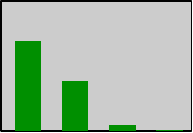}.

When looking at the failure cases (\ie non-EMs), the distribution of AST \emph{edit actions} that was needed to address the SATD (but that the model failed to reproduce) we found that out of the EMs: 32.01\% are \texttt{Deletions}, 54.10\% \texttt{Insertions}, 6.84\% \texttt{Moves}, and 7.05\% \texttt{Updates} \includegraphics[scale=0.17]{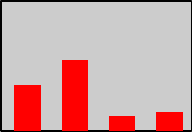}. Thus, are not the ``types'' of AST edits needed to address the SATD that discriminate what the model can or cannot do. Given this finding, we also computed the sheer number of AST \emph{edit actions} that were needed in EMs and wrong predictions to address the SATD. The achieved results, as expected, indicate that the model struggles in addressing SATD in need of a high number of AST changes to be repaid. Indeed, the median number of changes that were required to address the SATD instances that resulted in EMs is 6, as compared to the 20 of the wrong predictions (mean 9.8 \emph{vs} 38.1). 

\figref{fig:qualitative} illustrates four instances from our test set, including two for which M2$_{CC}$ was able to successfully address the SATD (\circled{1} and \circled{2}), and two for which it failed to pay back the TD (\circled{3} and \circled{4}). For the successful cases, the code on top shows the input method including the SATD, while the one at the bottom shows how the model addressed the SATD (changes highlighted by the yellow boxes). For the failure cases, we also report the expected target from our dataset, namely the code showing how the developers actually addressed the SATD (changes highlighted with grey boxes).

In \circled{1} the SATD mentions ``\texttt{TODO Lowercase paradigm}''. To fulfill this requirement, the model performs two distinct AST edit actions, addressing the TD in the right location by injecting the \texttt{toLowerCase} invocation.
The example of scenario \circled{2} shows how the model addresses the SATD by replacing the \emph{reportDirectory} attribute in the current instance with \emph{resultsDirectory} through a code change that involves updating two AST nodes (\ie an \texttt{Update} edit) in the \emph{if} and \emph{else} statements.

In scenario \circled{3} the model fails to effectively handle the SATD comment (\texttt{TODO load from actual config}) by requiring modifications to the instantiation of \texttt{this.socketAddress}. The hard-coded \emph{IP address} and \emph{port} need to be replaced with values fetched using the \texttt{config} method. The implementation of these changes is unsuccessful, as the model outputs the same method ($M_{satd}$) with the SATD comment removed (gray box in \circled{3}). There are a total of 21 AST edit actions to be implemented to successfully address the SATD comment, including 18 \texttt{Insert} operations, two \texttt{Delete} operations, and one \texttt{Update} action. 
When considering scenario \circled{4}, the SATD left by the developer requires the complete implementation of the \emph{storeCredential} method. Nonetheless, the recommendation provided by the model does not address the SATD comment appropriately since it assumes the existence of a \emph{storeCredential} method that takes \emph{context, account}, and \emph{storage} as input parameters, failing to address the TD. A successful change would have required the addition of 19 new nodes (\ie \texttt{Insert}) to the AST of the \java method \emph{storeCredential}. 

%\vspace{-0.5cm}
\begin{resultbox}
	\textbf{Answer to RQ$_{2}$.} Seeding task-similar knowledge (\eg code changes) into a model pre-trained using self-supervised task models of code only slightly improves their performance. Even our best-performing model struggles in addressing SATD instances requiring a large number of AST edit actions.  
\end{resultbox}
\subsection{RQ$_{3}$: To what extent does the presence of ``context-specific knowledge'' help pre-trained models of code in the automated SATD repayment?}

By comparing the results in row 4 of \tabref{tab:perfect} (M$_{CC-Ablation}$) with those in row 3 (M$_{CC}$), we can observe the fundamental role played by the context-specific knowledge provided as input to the model (\ie the SATD comment) in automatically addressing TD.
Admitting TD through a comment aids the model to better perform for all considered beam sizes ($K$). For example, when focusing on a single candidate solution (\ie top-1), M3$_{CC-Ablation}$  can only achieve an EM in 0.9\% of cases, while  M2$_{CC}$ does it in 2.30\% of cases. When looking at higher values of $K$, the gap becomes even larger, up to a +3\% for $K=10$ (5.10\% vs. 8.10\%). 
McNemar's test indicates a significant difference ($p$-value < 0.05)  between M2$_{CC}$ and M3$_{CC-Ablation}$, with M2$_{CC}$ having 8 times higher odds ($OR$=8) to address SATD than M3$_{CC-Ablation}$. Instead, although the differences found by Wilcoxon signed-rank test for the CrystalBLEU are statistically significant, the effect size is \emph{negligible}. 
The obtained results further highlight the importance for developers to admit TD. In essence, SATD does not only serve as a trace for themselves and for other developers \cite{ZampettiFSP21}, but, also, as a way to better guide automated tools in addressing TD. Furthermore, this stresses the importance of recommending developers that TD should be admitted \cite{ZampettiNAKP17}.

\begin{resultbox}
	\textbf{Answer to RQ$_{3}$.} The availability of context-specific knowledge in the form of SATD comments enhances the performance of pre-trained models of code, allowing them to achieve a substantial increase in the percentage of automatically addressed TD. This is a further motivation for developers to admit TD in their source code.
\end{resultbox}
\vspace{-4mm}

\subsection{RQ$_{4}$: Are general-purpose large language models zero-shot learners for SATD repayment?} 

The last three rows of \tabref{tab:perfect} report the performances achieved by ChatGPT as \emph{zero-shot learner} for addressing SATD. Two findings emerge: (i) the usage of different templates to prompt  ChatGPT for the task of SATD repayment plays a crucial role and, (ii) the performances achieved when recommending one single candidate solution (top-1) are lower than DL-based techniques appositely fine-tuned (M1 and M2$_{CC}$) to pay back TD. 
As for the use of several prompt templates, it is important to note that designing templates showing the code first and the comment later (as did in M4$_{T3}$) strongly penalize the model, resulting in 0 EM and the lowest CrystalBLEU score across all treatments. 
Differently, providing the SATD comment first and the code including such a comment later helps ChatGPT in achieving better performances, with 1.18\% and 1.19\% of SATD comments successfully addressed, respectively for M4$_{T1}$ and M4$_{T2}$. 

McNemar's test on the top-1 recommendation indicates no significant differences ($p$-value > 0.05) between M4$_{T1}$ and M2$_{CC}$, as well as M4$_{T2}$ and M2$_{CC}$. However, there are statistically significant differences when comparing the CrystalBLEU distributions of the tested templates with M2$_{CC}$. In these instances, the effect is \emph{large} for M4$_{T1}$ and M4$_{T3}$, and small for M4$_{T2}$, where ChatGPT performed best.

\rev{Without knowing the details of ChatGPT's implementation, it is hard to speculate on the reasons behind such performances. Possibly, they could be related to the lack of a specific fine-tuning for the specific task, or also to the need to generate suitable prompts.}

\vspace{0.1cm}
\begin{resultbox}
	\textbf{Answer to RQ$_{4}$.} When used in a zero-shot setting, LLMs, and  ChatGPT in particular, exhibit sub-optimal performance compared to pre-trained models of code appositely tuned for the specific task of addressing SATD. Additionally, the use of well-crafted templates plays a crucial role for LLMs being used off-the-shelf.
\end{resultbox}

\section{Threats to Validity}
\label{sec:threats}
% !TEX root = main.tex

\textbf{Construct validity.} The main issue to be considered is whether the observed SATD removals are true positives. To mitigate this threat, we only considered cases in which the SATD comment contained well-recognized keywords, and we manually analyzed a sample to check for problematic cases. 

To avoid training our model on instances where the SATD was removed by chance, we excluded cases where the affected method was removed. We are, however, aware that the latter heuristics, while mitigating some threats to construct validity, could affect the study's generalizability.

\textbf{Internal validity.} As explained in \secref{sub:techniques}, we used the default settings of the employed language models. Better results could be achieved with proper hyperparameter tuning. \rev{The assumption we have made in Section III-B which links the removal of an SATD comment with the change performed within the same method is subject to imprecision. First, the commit could tangle the SATD repayment with other changes. Second, the comment removal may be out of sync with the source code change aimed at addressing the SATD.}

\textbf{Conclusion validity.} The comparison between different techniques is supported by suitable statistical procedures and effect size measures. Also, the results of multiple tests have been adjusted through Holm's procedure  \cite{Holm1979a}. \rev{We are aware that the performance of the studied approaches may change, and possibly improve, if experimenting with a larger dataset.}

\textbf{External validity.} Our study is limited in terms of programming language (Java) and, more important, the SATD removal dataset is based on 1,000 instances. \rev{As stated in \secref{sub:techniques}, we limit to addressing SATD that have been resolved within the same method, and to methods not longer than 1024 tokens. As the paper aims to set---within the employed generative models---an ``upper bound'' of the SATD repayment capability, we do not expect any better results for more complex and extensive changes.}
Better generalizability of our results would require studies on further and more diversified datasets. \rev{
In terms of considered models, our results are limited to the CodeT5-base~\cite{wang2021codet5}, as well as a zero-shot attempt done with ChatGPT~\cite{chatgpt}. Other models having a different size and architecture could, possibly, exhibit different results. However, in this circumstance, our interest was to mainly show the feasibility of SATD removal and, within the same model (CodeT5 in our case), the relative improvements with different levels of pre-training and fine-tuning. 
Moreover, we acknowledge that we have not experimented with edit-specific models, such as CoditT5 \cite{ZhangP0LG22}, and therefore further experimentation with such models would be desirable. At least, we partially mitigated this threat by experimenting in RQ$_2$ a pre-training with code changes. Moreover, as explained in Section II-C, our dataset mostly features removals and additions rather than updates and moves.}

\rev{Last, but not least, also the preliminary results with ChatGPT need to be confirmed or confuted with similar yet differently implemented tools, \eg Google Bard~\cite{bard}.}

\section{\rev{Discussion and Conclusion}}
\label{sec:conclusion}
% !TEX root = main.tex

In this paper, we investigated the use of Deep Learning (DL) models for automatically addressing technical debt (TD). To train the models, we leveraged 5,039 instances of SATD removals mined using an existing tool \cite{alomar2022satdbailiff}. Such a small number of training instances made the pre-training of the model absolutely necessary to achieve an automated fixing of SATD instances. Nevertheless, even the best-performing model we experimented with can automatically repay SATD only in a minority of cases (2\% to 8\%). The complexity of such a task has also been confirmed by the results achieved exploiting the state-of-the-art LLM (\ie ChatGPT \cite{chatgpt}), that under-performed if compared to the specialized models we tested.
\rev{So far, the use of generative deep learning models has been successful for several code tasks that require either generation of new code, or the change of existing one. Some (non-exhaustive) examples of achieved performances are on the order of $\sim$14\% for bug fixing (Wang \etal \cite{wang2021codet5}), of $\sim$5\% for generating a reviewed version of existing source code (Tufano \etal \cite{tufano2022using}), and $\sim$23\% for generating code blocks (Ciniselli \etal \cite{ciniselli2021empirical}).} 
\rev{As far as SATD repayment is concerned, we have observed both positive results and negative results. On the positive side, results are in line with some existing approaches for automated bug fixing \cite{tufano2019empirical}.}

\rev{On the negative side:
	\begin{compactenum}
		\item The level of performances achieved so far would still limit the applicability of the proposed approach in real practice. The latter may also possibly require some explanation/rationale of the changes to be performed, \eg telling to the developer that the change being done is enacting a refactoring action, fixing a bug, improving the code readability, etc.
		\item Such results have been obtained under the assumption that the change concern a single method, with a limited maximum size of the considered methods, and assuming that the removal of a SATD comment would correspond to changes aimed at addressing it. 
		\item Although the proposed approach was able to correctly recommend instances of SATD repayment involving the addition of an AST (over 33\% of our exact matches), the approach may fail to introduce a totally new, unseen piece of feature, as well as it is unlikely to be able to perform changes such as API upgrade/replacements. 
\end{compactenum}}

\rev{The aforementioned limitations greatly stimulate future research in this area. First and foremost, although in RQ$_2$ we have performed a large training of CodeT5 on a dataset of code changes, it would be worthwhile to experiment with models more specifically suited for code edits, such as CoditT5 \cite{ZhangP0LG22}.}

\rev{Second, although RQ$_4$ has shown that a zero-shot instance of ChatGPT fails to support the SATD repayment task, other investigations with LLMs are worthwhile, as those might be able to help in supporting larger change tasks. They can go in the direction of prompt engineering, as well as experimenting with other LLMs.}
\rev{Third, given the variety of the SATD nature, it may be worthwhile to pursue the development of eclectic approaches, that first classify the type of needed change (\eg along the line of what SARDELE \cite{ZampettiSP20} does) and then employing different models or even different approaches for each of them.} 
Our future work targets (i) the possibility to exploit LLMs in a few-shot learning scenario (rather than the zero-shot we experimented with), and (ii) the definition of different mining pipelines which could help enlarging the SATD removal dataset, possibly boosting performance.

\section{Data Availability}
\label{sec:replication}
% !TEX root = main.tex

Our replication package \cite{replication} includes all code and data employed in our research. This comprises the datasets needed for training and testing the models, the code to reproduce the experiments (\ie models' training and inference), the predictions generated in each setting, and the \texttt{R} scripts employed for performing statistical analysis.

\section{Acknowledgments}
\label{sec:ack}
This project has received funding from the European Research Council (ERC) under the European Union's Horizon 2020 research and innovation programme (grant agreement No. 851720).
Di Penta acknowledges the Italian “PRIN 2021” project EMELIOT ``Engineered MachinE Learning-intensive IoT systems.''

\balance
\bibliographystyle{IEEEtran}
\bibliography{main}

\end{document}